\documentclass[5p,times]{elsarticle}
\usepackage{graphicx} 
\usepackage{multirow}%
\usepackage{amsthm,amssymb,amsfonts}
\usepackage{amsthm}%
\usepackage{mathrsfs}%
\usepackage[title]{appendix}%
\usepackage{xcolor}%
\usepackage{textcomp}%
\usepackage{manyfoot}%
\usepackage{booktabs}%
\usepackage{algorithm}%
\usepackage{algorithmicx}%
\usepackage{algpseudocode}%
\usepackage{listings}%
\usepackage{lscape}
\usepackage{endnotes}
\usepackage{adjustbox}     
\usepackage{url}
\usepackage{mathtools}
\theoremstyle{definition}
\newtheorem{definition}{Definition}[section]
\usepackage[flushleft]{threeparttable}
\begin{document}

\begin{frontmatter}



\title{A Quantization-based Technique for Privacy Preserving Distributed Learning}


\author[inst1]{Maurizio Colombo}
\ead{maurizio.colombo@ku.ac.ae}
\author[inst2]{Rasool Asal}
\ead{rasal@ud.ac.ae}
\author[inst3,inst4]{Ernesto Damiani\corref{cor1}}
\ead{ernesto.damiani@ku.ac.ae}
\author[inst1,inst5]{Lamees Mahmoud AlQassem}
\ead{lamees.alqassem@ku.ac.ae}
\author[inst3]{Al Anoud Almemari}
\ead{100041410@ku.ac.ae}
\author[inst3]{Yousof Alhammadi}
\ead{yousof.alhammadi@ku.ac.ae}

\affiliation[inst1]{organization={EBTIC, Khalifa University},
            addressline={P.O. Box: 127788}, 
            city={Abu Dhabi},
            country={UAE}}
\affiliation[inst2]{organization={College of Engineering and IT, University of Dubai},
            addressline={Academic City, Emirates Road}, 
            city={Dubai},
            country={UAE}}
\affiliation[inst3]{organization={C2PS, Khalifa University},
            addressline={P.O. Box: 127788}, 
            city={Abu Dhabi},
            country={UAE}}
\affiliation[inst4]{organization={Computer Science Department, Univ. degli Studi di Milano},
            addressline={via Celoria 18}, 
            city={Milan},
            country={Italy}}
\affiliation[inst5]{organization={Department of Computer Science, Khalifa University},
            addressline={P.O. Box: 127788}, 
            city={Abu Dhabi},
            country={UAE}}
 \cortext[cor1]{C2PS, Khalifa University, P.O. Box: 127788, Abu Dhabi, UAE, ernesto.damiani@ku.ac.ae, phone +971 2 312 4411}
 
\begin{abstract}
The massive deployment of Machine Learning (ML) models  raises serious concerns about data protection. Privacy-enhancing technologies (PETs) offer a promising first step, but hard challenges persist in achieving confidentiality and differential privacy in distributed learning. In this paper, we describe a novel, regulation-compliant data protection technique for the distributed training of ML models, applicable throughout the ML life cycle regardless of the underlying ML architecture. Designed from the data owner's perspective, our method protects both training data and ML model parameters by employing a  protocol based on a quantized multi-hash data representation \emph{Hash-Comb} combined with randomization. The hyper-parameters of our scheme can be shared using standard Secure Multi-Party computation protocols.
Our experimental results demonstrate the robustness and accuracy-preserving properties of our approach.
\end{abstract}




\end{frontmatter}



\section{Introduction}

The ongoing massive deployment of Machine Learning (ML) models involves collection, storage and exchange of huge amounts of training data, which raised serious concerns about data protection and attracted the attention of regulatory bodies worldwide \cite{pugliese2021machine}. To name but one, the European General Data Protection Regulation (GDPR) imposes obligations onto organizations anywhere, so long as they target or collect data related to people in the EU.  The regulation mentions the use of \emph{Privacy-enhancing technologies} (PETs) to protect personal data \cite{el2024preserving}. In the last few years, many attempts have been made \cite{jegorova2022survey} to develop specific PETs to achieve confidentiality and privacy in distributed training of Machine Learning (ML) models. 

Nevertheless, there are still many privacy and confidentiality threats affecting processing and communication of training data and ML models' parameters \cite{caroline2020artificial}. 
While  methodologies for modeling, detecting and alleviating such threats are still being developed \cite{mauri2021stride}, organizations are required to satisfy privacy regulations when carrying out distributed learning, especially when personal data are involved \cite{caroline2021securing}.

In this paper, we describe a novel data protection technique designed from the viewpoint of the data owners who hold  training data sets,  ML model parameters, or both, and want to protect them independently of the specific ML architectures and training algorithms. 
These data owners need a way to make sure that the PET choice and configuration they use are both effective and regulation-compliant.
To tackle this challenge, we combine randomization of quantization as proposed in \cite{youn2023randomized} with \emph{Hash-Comb}, our own quantized multi-hash data representation \cite{almahmoud2022hash} to alleviate privacy and confidentiality threats to distributed Machine Learning (ML) while guaranteeing regulatory compliance.
The major contribution of this paper is threefold:
\begin{itemize}
\item An entirely novel, efficient and easy-to-implement method to achieve the desired level of differential privacy of ML models' parameters and data exchanged in distributed training by adding randomized quantization noise.
\item A distributed protocol based on secret sharing, integrating our method with existing frameworks for federated learning.
\item Integrated support for parameter and data confidentiality, both in-transit and at-rest, using only standard hashing function that comply with regulations.
\end{itemize}
Our experimentation on multiple datasets shows that our method provides faster training convergence and better accuracy of the final ML models compared to benchmark techniques.

\subsection{Background and Problem Statement}

To introduce the main privacy and confidentiality threats to ML models' training, let us consider a typical ML task: classifying the items of a set $DS$ into classes of interest belonging to a finite set $C = (C_1, ...,C_n)$ \cite{cimato2018some}. 

As we do not have access to the entire domain $DS$, we use a \emph{sample} $S \subset DS$ of the domain and tabulate a partial classification function $f() : S \rightarrow C$, obtaining a subset of $DS \times C$, which, with a slight abuse of notation, we call $f$. Each entry $e$ of $f$ has the form $\{s,C_i\}$ where $s\in S$ and $C_i \in C$. We can then use $f$ as a \emph{training set} to train a supervised ML model that computes an estimator function $F : DS \rightarrow C$, hopefully coinciding with $f$ at (most of) the points belonging to $S$.

The model $F$ can be either deployed into production to classify elements of $DS$ as they become available (the so-called \emph{inference} step); alternatively, one can repeat the process, collecting $k$ samples $S_1,...,S_k$ in $DS$ and training the corresponding ML models $F_1,...F_k$. These estimators can then be merged to obtain a \emph{federated} global estimator $F_G$.

This procedure is subject to two potential threats: a \emph{threat to privacy} regarding the entries of the training set $f$, if $f$ can be reverse-engineered from $F$ (or $F_G$), and a \emph{threat to confidentiality} regarding the parameters of the ML models $F_i, i=1,..,k$ when they are communicated to the central server in charge of merging them into $F_G$.
The privacy violation will happen if, by observing the outcome of $F$, an attacker can reconstruct one or more entries of $f$. To alleviate the privacy threat, we can pursue a \emph{non-disclosure} goal: given any entry $e \in f$, any observer of the execution of $F$ should be able to infer the same information about $f$ as by observing any other model $F'$ obtained using the training set $f - {e} + {r}$, where $r$ is a random entry. 

Two decades ago, Cynthia Dwork proposed the notion of \emph{differential privacy} (DP) \cite{dwork2006differential} to provide a quantitative measure of the degree of satisfaction of the above goal after any data transformation. Other seminal papers \cite{dwork2008differential}, \cite{ghosh2009universally}, \cite{chen2011publishing} have shown that it is possible to add noise to data sets and alleviate the threat to privacy while preserving some measure of accuracy of any ML model trained on them. 

\subsection{Differential Privacy}
\label{DP}
Formally, a ML model $F$ trained on $f$ guarantees a level $\epsilon$ of \emph{differential privacy} (DP) if, for all possible training sets $f'$ differing in a single value from $f$, for all outputs $C_i \in C$ and for all $x \in DS$ we have:
\begin{equation}
(1 - \epsilon) \leq {{Pr(F(x) \in C_i)} \over
{Pr(F'(x) \in C_i)}} 
\leq (1 + \epsilon)
\label{eqdp}
\end{equation}
where $F'$ is (the same) model trained over $f'$. 

A popular technique for achieving the desired level of differential privacy is \emph{non-interactive} DP \cite{leoni2012non}, which consists in adding random noise to an existing training $f$ to obtain an \emph{adjacent} training set $f'$.  A convenient probability density function (PDF) for such noise is the Laplace distribution \cite{Laplace}:
\begin{equation}
p(z) = e^{-{|z|\over{\sigma}}} = e^{-|z|\epsilon} 
\end{equation}
This distribution is “concentrated around the truth”: after generating an adjacent data set by noise addition, the probability that an entry $e$ in it will fall $z$ units from the true value drops off exponentially with $z$, with a progression given by the $\epsilon$ parameter. In other words, a lower $\epsilon$ increases the probability of values far from the truth, achieving stronger privacy.
Besides Laplace, other distributions have been used for DP, including the Gaussian \cite{dong2022gaussian}, the geometric ] \cite{ghosh2009universally}, and the exponential one \cite{ExponentialDP}.
Due to the noise, DP introduces some uncertainty: an ML model trained of a training set $f'$ adjacent to $f$ will no longer compute $F$ but another function  $F'$, potentially less accurate in production.
The need for balancing data privacy and accuracy brought to a more general definition of \emph{interactive} differential privacy (NIDP) \cite{lyu2022composition}, also known as ($\epsilon, \delta$)-DP. Again, ($\epsilon, \delta$)-DP is based on the concept of adjacent data sets, i.e. data sets that differ in just one entry. It can be formally defined as follows \cite{DP_Defn}:
\begin{definition}
Let $S$ and $S'$ be two adjacent data sets. Then, a randomized function $F:\chi \rightarrow C$ satisfies ($\epsilon, \delta$)-DP if for any $S, S'\in \chi $ and any subset of outputs ($O \subseteq C$):
\begin{equation}
Pr(F(S)\in O) \leq e^{\epsilon}Pr(F(S')\in O)+\delta
\end{equation}
where:
 \begin{itemize}
     \item $Pr(F(S)\in O)$ is the probability of model $F$ to give an output in $O$ for an input in $S$.
     \item $Pr(F(S')\in O)$: probability of model $F$ to give an output in $O$ for input in $S'$.
     \item $\epsilon$ is a non-negative parameter, also known as \emph{privacy} budget, that quantifies the level of privacy protection. 
     \item $\delta$ is a non-negative parameter that represents the \emph{failure probability} of not holding the privacy guarantee. The probability of privacy breach is bounded by $\delta$. 
 \end{itemize}
\end{definition}
Values of $\delta$ and $\epsilon$ can be adjusted according to requirements and even modified interactively \cite{lyu2022composition}.
In 2017, Ilya Mironov introduced the notion of \emph{R{\'e}nyi differential privacy} (RDP) \cite{mironov2017renyi}. RDP requires the \emph{divergence} between the distributions of the original and the noisy samples $S$ and $S'$ to be properly bounded. 
We provide a formal definition of R{\'e}nyi divergence in Section \ref{RDP}. A key property of RDP is that a mechanism which provides RDP also provides classic DP. \\

The remainder of this article is organized as follows: In Section \ref{RelatedWork}, we review related work on quantization-based privacy in distributed learning. In Section \ref{ADPQ}, we present our own approach.
In Section \ref{MHR}, we show how standard hashing can be used to turn the quantized representation into our multi-hash encoding, \emph{Hash-Comb} \cite{almahmoud2022hash}.
In Section \ref{NP}, we outline how the hyper-parameters of our representation scheme can be securely shared using \emph{Multi-Party Computation} (MPC).
In Section \ref{THC}, we discuss centralized and federated training of ML models on randomized \emph{Hash-Comb} data. In the context of Federated Learning, we show that our method preserves the confidentiality of local models' parameters while allowing for the computation of an accurate global model $F_G$.
In Section \ref{E}, we carry out an experimental comparison between our approach and other techniques for achieving DP by noise addition. Finally, in Section \ref{C}, we draw our conclusions and highlight some future challenges and application scenarios.

\section{Related Work}\label{RelatedWork}
Much research has targeted communication efficiency and privacy preservation in distributed learning \cite{Quan_LDP, yan2023killing,cpSGD, yan2023layered, youn2023randomized}. 
Already early research recognized that privacy can be improved by quantization of model updates, particularly gradients \cite{cpSGD,Quan_LDP}. Pioneering work \cite{cpSGD} presented a differentially private distributed stochastic gradient descent (DP-SGD) algorithm  where  quantization is randomized via a binomial mechanism, serving as a discrete counterpart to the Laplacian or Gaussian mechanisms outlined in Section \ref{DP}. Zong et al. \cite{Quan_LDP} proposed \emph{Universal Vector Quantization for Federated Learning (FL}) with Local Differential Privacy (LDP). In this approach, FL clients achieve LDP by introducing Gaussian noise to their model updates. Then, the noisy updates are quantized and encoded into digital code-words to be transmitted to the server for aggregation. 

This "quantization-then-noise" family of approaches showed some limitations in accuracy and efficiency, leading to biased estimation due to modular clipping. The alternative method is swapping the order of the operation, adding noise to  quantized parameters. For instance, Yan et al. \cite{yan2023killing} add binomial noise to uniformly quantized gradients. Their approach attains the desired level of differential privacy with only a slight reduction in communication efficiency.

Recent research is based on the notion that quantization alone may achieve the desired privacy level. Several research efforts, such as \cite{youn2023randomized,yan2023layered}, successfully leveraged quantization to simultaneously optimize communication efficiency and preserve privacy in distributed learning. These works demonstrate that the perturbation introduced by the compression process can be translated into quantifiable noise. The work closest to ours \cite{youn2023randomized}, achieves R{\'e}nyi differential privacy (RDP) through a two-tiered randomization process. Initially, it randomly selects a subset of feasible quantization levels, followed by a randomized rounding procedure utilizing these discrete levels. 
A more recent work \cite{yan2023layered} introduces Gau-LRQ (Gaussian Layered Randomized Quantization), leveraging layered multi-shift couplers to compress gradients while shaping the error towards a Gaussian distribution, thereby ensuring client-level differential privacy (CLDP).  Gau-LRQ is implemented within the distributed stochastic gradient descent (SGD) framework and is evaluated on various datasets, including MNIST, CIFAR-10, and CIFAR-100. The results demonstrate better performance compared to the approach proposed in \cite{yan2023killing} and to a baseline that uses Gaussian DP only, without quantization.



All methods reviewed above enhance privacy and communication efficiency in the context of gradient-based large-scale distributed learning. We follow a different approach, pursuing a random quantization technique applicable to any value, be it training data or model updates. Our approach can achieve differential privacy in a general setting without imposing constraints on the model parameters or integrating other privacy techniques. Our approach also achieves efficient transmission due to the fixed message size exchanged between the server and client. A summary of our technique's features and a comparison to the ones of other approaches are presented in Table \ref{Table:LR}.

\begin{table*}[]
\centering
\caption{Quantization-based Approaches Comparison}
\label{Table:LR}
\resizebox{\textwidth}{.09\textwidth}{%
\begin{tabular}{|c|c|c|c|c|c|}
\hline
\textbf{Ref} & \textbf{Random Quantization} & \textbf{Extra Privacy technique} & \textbf{DP-Noise Addition} & \textbf{differential privacy guarantee} & \textbf{Application} \\ \hline
\cite{cpSGD} & $\times$ & $\checkmark$ & Gaussian Noise & $\checkmark$ & gradient-based large-scale distributed learning \\ \hline
\cite{Quan_LDP} & $\times$ & $\checkmark$ & Gaussian Noise & $\checkmark$ & gradient-based FL model \\ \hline
\cite{yan2023killing} & $\times$ & $\checkmark$ & Binomial noise & $\checkmark$ & gradient-based large-scale distributed learning \\ \hline
\cite{youn2023randomized} & $\checkmark$ & $\times$ & not needed & $\checkmark$ & General \\ \hline
\cite{yan2023layered} & $\checkmark$ & $\times$ & not needed & $\checkmark$ & gradient-based large-scale distributed learning \\ \hline
Our Work & $\checkmark$ & $\times$ & not needed & $\checkmark$ & General \\ \hline
\end{tabular}%
}
\end{table*}

\section{Achieving Differential Privacy via Quantization}
\label{ADPQ}

In this section, for the sake of simplicity, we will assume that each data point in $DS$ is a scalar $x$; our discussion is readily extendable to any vector data space ${\bf x}=x_1,...,x_n$. The quantization of samples in each sampling domain $S$ involves three steps:
\begin{itemize} 
\item Negotiating the overall quantization range and the number $L$ of possible quantizations. 
\item Performing a \emph{randomized selection} of a number $l < L$ of the quantizations to apply to the sampled $x$ values in $S$. 
\item Computing the quantized image of the data points in the sample $S$. Each point in $S$ corresponds to $l$ quantized values. 
\end{itemize}
 We  describe each step in  detail below.

\begin{enumerate}
\item {\bf Range Enlargement} This step establishes the range of values in the quantized image of $S$.
We subtract a parameter $\Delta$ from $x_{min}$, the smallest $x \in S$ and adding $\Delta$ to $x_{max}$, the largest $x \in S$. The enlargement relative size is expressed by the ratio $x_{max} \over {\Delta}$.
The enlargement of the value range is necessary to achieve privacy: if we used for the quantized image of $S$ the same range of $S$, the quantization outputs for $x_{max}$ and $x_{min}$ would always end up in the top and bottom channels of each quantization, leaking information about $x$. Within this augmented range we establish $L$ quantization levels, respectively involving $m_1,..,m_i,..,m_L$ evenly spaced quantization channels $(C(0), C(1), · · · , C(m_i - 1))$.
\item {\bf Sub-sampling of quantizations} This step randomly sub-samples $l$ of the possible $L$ quantization levels, including each level of quantization in our encoding with a \emph{selection probability} $p>0$. 
The value $p=0.5$ corresponds to standard fair coin randomization. As we will see in Section \ref{MHR}, different values of $p$ can be used to achieve the desired randomization (i.e., the target value of $l$) with a limited number of extractions, neutralizing selection bias \cite{frane1998method}.
\item {\bf Quantized image computation} In this step, we compute the multi-value quantized image of $S$. Each point $x \in S$ generates a list of values, corresponding to the channel where $x$ falls in each of the $l$ sub-sampled quantizations. At each level, an approximation of the value of $x$ can be extracted by estimating $x$ as the mid-point of the quantization channel it fell in. Section\ref{MHR} includes a worked-out example of this computation.
\end{enumerate}

\subsection{R{\'e}nyi Differential Privacy}
\label{RDP}

We now discuss how our scheme achieves R{\'e}nyi differential privacy. We start with defining \emph{divergence}.
Informally, divergence is a binary function which establishes the separation from one probability distribution to another. The simplest divergence is squared Euclidean distance. Other classic divergences like \emph{Kullback–Leibler} (KL) divergence\cite{agrawal2019unifying} can be seen as generalizations of it.\\
Here, we are interested in the \emph{R{\'e}nyi divergence} of positive order $\alpha \neq 1$ between a discrete probability distribution $P = (p_1,...,p_n)$ and another distribution $Q = (q_1, . . . , q_n)$. It can be written as follows \cite{van2014renyi}:
\begin{equation}
D_\alpha(P\vert \vert Q) = {{1}\over{\alpha-1}} ln \sum_{i=1}^{n}p_i^{\alpha} q_i^{1-\alpha} 
\end{equation}
where, for $\alpha > 1$, we read $q^{1-\alpha}$ as $q^{{\alpha}-{1}}$ and adopt the conventions that $0/0 = 1$ and $x/0 = \infty$ for $x > 0$.



We recall the relationship between the R{\'e}nyi divergence with $\alpha = \infty$ and $\epsilon$-differential privacy (Section \ref{DP}, Eq.\ref{eqdp}). 
Any randomized data representation mechanism is \emph{differentially private} if and only if its distribution over any two adjacent data sets $S$ and $S'$ satisfies the following equation:
\begin{equation}
    D_\infty(f(S) \vert\vert f(S') )\leq \epsilon
\end{equation}
 To state this property for the distributions of any pair of adjacent samples in the multi-quantization procedure of Section \ref{MHR}, we consider a point $x_j \in S$ and a replacement value $x'_j$ (so that $S$ and $S'=S-\{x_j\}+\{x'_j\}$ are adjacent).
Let us call $P$ and $Q$ the probability distributions of the (multi-level) quantized representations of $S$ and $S'$ across the same random selection of the quantization levels.
We consider the finest grained quantization in our representation, and map each data point $x_i$ ($i=1,...,n$) to a quantized value $t_h$ for $h=1,..c$ where $c$ is the number of quantization bins. The $c$-dimensional distribution of values $t_h$ is itself a quantization of the $n$-dimensional discrete probability distribution of the data values, with a given number $c$ of supporting points \cite{roychowdhury2022quantization}, coinciding with the bins' midpoints.
Let $p_h$ ($q_h$) be the number of occurrences of each value $t_h$ divided by the size $\lvert S \rvert$ of $S$ (or $S'$). In other words $p_h = {{\#(t_h)} \over {\vert S \vert}}$. 
As $S$ and $S'$ differ only in $x_j$ being replaced by $x'_j$, $P$ and $Q$ contain the same quantization, so $p_h = q_h$  for $h<>r,s$ where the $r$ and $s$ are the (possibly coinciding) quantization bins where $x_j$ and $x'_j$ have fallen. We get:
    \begin{equation} 
    D_\alpha(P||Q) = {{1}\over{\alpha-1}} ln (\sum_{h=1}^{k}p_h^{\alpha} q_i^{\alpha-1}) =
        \end{equation} 
        \begin{equation}   
        {{1}\over{\alpha-1}} ln (\sum_{h<>r,s}^{n}p_h^{2\alpha-1} +p_r^{\alpha} q_s^{\alpha-1})
        \label{capx}
         \end{equation}

where $p_r$ and $q_s$ are the frequencies (i.e., the normalized occurrences) of the quantization bins where $x_j$ and $x'_j$ have fallen. We can now cap the value of our Equation \ref{capx} as follows:
         \begin {equation}        
         {{1}\over{\alpha-1}} ln (n \  p_{max}^{2\alpha-1}) =
             \end{equation} 
             \begin{equation}
             {{1}\over{\alpha-1}} (ln (n)+ (2\alpha-1) ln (p_{max})) = 
             \end{equation} 
            \begin{equation}
             {{1} \over{\alpha-1}} ln (n) + {{2 \alpha-1} \over{\alpha-1}} ln \ (p_{max})
             \end{equation}
\label{cap1}
where $p_{max}$ is the largest value among all $p_h$.
For $\alpha \rightarrow \infty$ we get:
\begin{equation}
    D_\infty(P||Q) \leq \epsilon = 2 \  ln(p_{max})
\label{cap2}
\end{equation}

Equation \ref{cap2} shows that, for any Hash-Comb, the R{\'e}nyi divergence between the probability distributions of two adjacent samples' quantizations is bounded by a constant, which can be tuned by adjusting the channel width of the finest-grained quantization.
Therefore, our data representation is R{\'e}nyi-differentially private and also differentially private in the classic sense \cite{youn2023randomized}.
Our Hash-Comb representation provides a small set of hyper-parameters whose setting can quantify the leakage of private information within the framework of R{\'e}nyi Differential Privacy\cite{liu2019private}. Such hyper-parameters can be chosen at each training run. 
Literature studies \cite{papernot2021hyperparameter} show that the reference RDP range is from $4$ to $7$. Using a biased coin with tunable selection probability $q$ and ${x_{max}\over{\Delta}}=2$, we can achieve any value in this reference range of RDP by using a relatively small number ($l=4$ to $l=8$) of quantization levels.

\section{Multi-Hash Representation } 
\label{MHR}


\
In this Section we compute our multi-quantization encoding for the parameters of a ML model~\cite{almahmoud2022hash}. Our method consists of hashing the quantized values, turning each multi-quantized value into a multi-hash representation. We use regulation-compliant, small-footprint hashes suitable for efficient communication. 

Encoding is performed at each round of the FL protocol as detailed in Algorithm~\ref{alg:HCAlg}. At each round, the local model parameters are encoded before being submitted to the central unit for averaging.
The Hash-Comb algorithm uses \emph{k} quantizers whose quantization level $1\leq l \leq k$ is defined by $2^l$ channels. Once established the channel where it falls at each level, the encoding value \textit{w} is rounded to the value representing the midpoint of the channel itself, providing an approximation up to $S(1/2)^{l+1}$, where \textit{S} is the size of our initial data space. 

Finally, the encoding for \textit{w} is obtained with the concatenation of $k$ hashes, each representing a specific branch of the binary graph in which the root is $S$ and each node in the path is the hashed value of the channel containing \textit{w} at every quantization level.  
The sequence of (hashed) channels captures different granularity of distance information from $S$. The last channel represents the finest-grained approximation for \textit{w} in our encoding, and represents the output of the decoding process. 

At this point, we need to identify the most suitable value for \emph{k}. Proceeding empirically, we found that encoding using $8$ quantization levels does not affect the outcome of model training, and in some cases even results in enhanced performance (Section~\ref{E}). This "magic number" of eight quantizers is also consistent with literature results (Section~\ref{RDP}) about the quantizers required to achieve RDP.
Another requisite to address is the one regarding sub-sampling (Section~\ref{ADPQ}), which dictates a theoretical maximum $L$ of quantization levels. Once more, we proceeded empirically by assuming a conservative scenario with an enlarged range $[-1, +1]$ and $L=16$. In this case any decoded \textit{w} would have an approximation error up to $\pm1.52587890625e-05$, which can be considered negligible for the training of the model. Adding more levels to the algorithm would only exponentially increase the computational workload without additional benefits.

\begin{figure}
  \centering
  \includegraphics[scale=0.5]{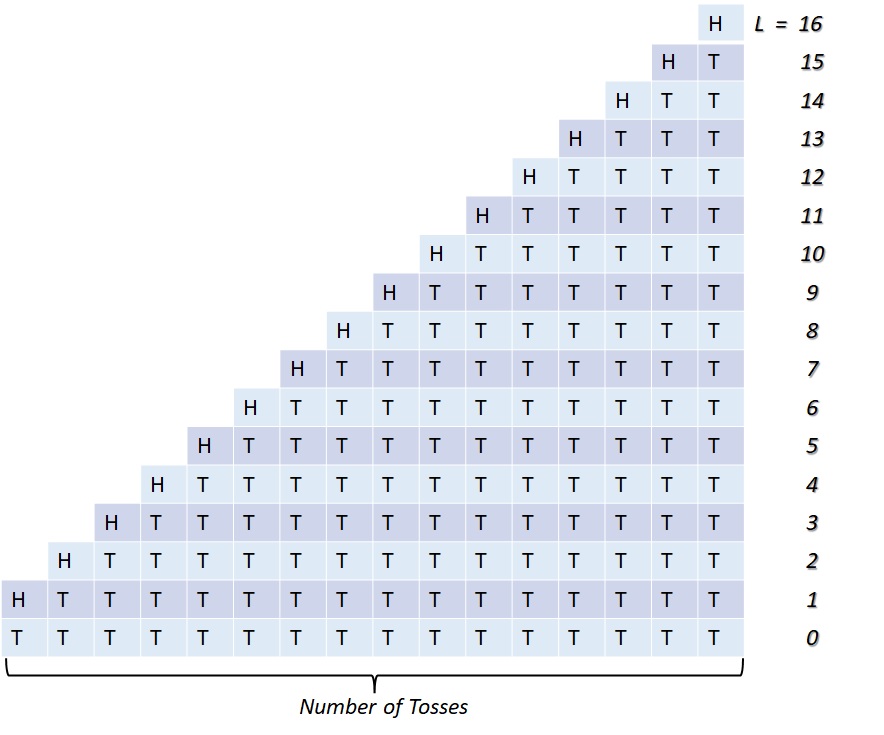}
  \caption{Hash-Comb levels as numbers of consecutive tosses of a biased coin. }
  \label{fig:tosses}
\end{figure}

\begin{figure*}
  \includegraphics[width=\linewidth]{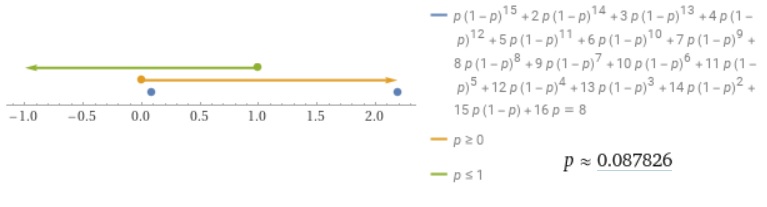}
  \caption{Solution for p when $\overline{k} = 8$.}
  \label{fig:Q8}
\end{figure*}

To sub-sample, we assign different levels of quantization to each parameter \textit{w} while maintaining an average $\overline{k}$ equals to $8$, as discussed above. The randomization factor in selecting the level is modelled as tossing a coin for $16$ times. The last toss in which \textit{Head} occurs indicates the quantization level for the current \textit{w}, as shown in Figure~\ref{fig:tosses}. 
A \emph{fair coin} where each outcome has probability $p = 0.5$ is unsuitable to our scheme. It results in $\overline{k} = 15$, as easily demonstrable through the formula~(\ref{eq:fairtosses}), which does not guarantee the desired privacy. To balance the average quantization across all weights and provide the necessary noise we use a biased coin with $Pr(H)=p$ and $Pr(T)= (1 - p)$ guaranteeing $\overline{k} = 8$. The solution $p = 0.087826$ was found by solving the linear system~(\ref{eq:biastosses}) shown in Figure~\ref{fig:Q8}. 

\begin{algorithm}
\caption{The algorithm computing Hash-Combs on each node at each communication round}
\label{alg:HCAlg}
\begin{algorithmic}
\Require $N \geq 1, R\geq 1$ 
\Ensure $k_i \in \mathbb{Z}, \quad  1 \leq k_i \leq L$
\Procedure{HashCombAlg}{$N,R$} \Comment{number of nodes and rounds}
\State $r\gets 0$
      \While{$r\not=R$}    
        \State $H\gets \{\}$ \Comment{initialize the communications}
        \For{\texttt{all $i \in N$}} \Comment{this is performed locally on each node $i$}
            \State \texttt{<compute $W^i$>} \Comment{$W$ result of $n$-cycles of optimization algorithm}
            \For{\texttt{all $w^i_j \in W^i$}}
                \State $k_j\gets$ \texttt{<randomly select $k$>}
                \State \texttt{$h_j = HC_{k_j}(w^i_j)$} \Comment{HC with a quantization layer $k$}
                \State $H^i\gets$ add($h_j$)
            \EndFor
            \State $H\gets$ add($H^i$) \Comment{add the locally encoded weights to the queue}
        \EndFor
        \State \textbf{return} $H$\Comment{all encoded parameters sent for FedAVG}
        \State $r = r+1$
      \EndWhile\label{HashCombLoop}      
    \EndProcedure
\end{algorithmic}
\end{algorithm}


         \begin {equation}        
            {\overline{k}} = \sum_{i=0}^L (L-i) \cdot Pr(k_{L-i})
            \label{eq:fairtosses}
         \end{equation} 
         \begin{equation}
           {\sum_{i=0}^L (L-i) \cdot p(1-p)^i = 8, \quad  0 \leq p \leq 1}
           \label{eq:biastosses}
         \end{equation} 

\section{The Negotiation Protocol}  
\label{NP}
In our scheme all participants must securely share the initial parameters of the quantization procedure, namely the $l < L$ selected quantization levels to be used and the enlarged ranges $[c_{min}, c_{max}]$, with $c_{min} = (x_{min} -\Delta$) and $c_{max} = (x_{max} +\Delta$).
Many different \emph{Multi-Party Computation} (MPC) protocols can be used for achieving common knowledge of these parameters without sharing the local feature values or ranges  \cite{cryptoeprint:2020/300}. Below, we describe a secure sharing protocol under the reasonable assumption of a  majority of honest participants.

\subsection{Secret Sharing}
Our protocol utilises the classic technique of \emph{Shamir secret sharing} \cite{shamir1979share} to share a secret $s$ amongst $n$ parties, so that any subset of $t + 1$ or more of the parties can reconstruct the secret, yet no subset of $t$ or fewer parties can learn anything about the secret. 
Shamir’s secret sharing scheme relies on the fact that for any for $t+1$ points on the two dimensional
plane $(x_1,y_1),...,(x_{t+1},y_{t+1})$ with unique $x_i$, there exists a unique polynomial $q(x)$ of degree
at most $t$ such that $q(x_i) = y_i$ for every $i$, and that it is possible to efficiently reconstruct
the polynomial $q(x)$, or any specific point on it\footnote{All computations in this section are in a finite field $Z_p$, for any prime number $p > n$}. 
In order to share a secret $s$, for example its local $x_{min}$, the participant chooses a random polynomial $q(x)$ of degree at most $t$ under the constraint that $q(0) = s$. 

In practice, each participant sets $a_0 = s$, chooses random coefficients $a_1,...,a_t \in Z_p$,and sets $q(x)= \sum_{i=0}a_i·x^i$. 
Then, for every $i=1,...,n$ the participant provides the $i$-th party with the share $y_i = q(i)$. Note that $p > n$, so different shares can be given to each party.
Reconstruction by a subset of any $t$ parties works by interpolating the polynomial to derive $q(x)$ and then compute $s = q(0)$. 

Although $t + 1$ parties can completely recover $s$, any subset of $t$ or fewer parties cannot learn anything about $s$, as they do not have enough points on the polynomial. Since the polynomial is random, all polynomials are equally likely, and  all values of $s \in Z_p$ are equally likely.

\subsection{The MPC protocol}
We are now ready to outline the steps of our protocol. 

\begin{enumerate}
\item {\bf Coordinator election}. The first conceptual step in our protocol is selecting the coordinating node who will be in charge of deciding the enlargement parameter $\Delta$. In practice, this step can be merged with the next one, as the node in charge of announcing $\Delta$ can simply be the one who shared the largest local $x_{max}$.
\item {\bf Local range sharing}: In this step, each party $i$ shares its local $x_{min_i}$ and $x_{max_i}$ with the other parties, using Shamir’s secret sharing. 
\item {\bf Quantization set-up} The coordinator decides the number of quantization levels $L$ and carries out the random extraction of the $l < L$ quantization levels to be used in the training.
\item {\bf Hyper-parameter sharing} The coordinator announces the $x_{min}$, $x_{max}$, $\Delta$ parameters and the $l$ extracted quantization levels to the other parties, using Shamir’s secret sharing. 
\end{enumerate}
The protocol must be executed before distributed training can begin. It is secure for semi-honest adversaries, as long as less than $n/2$ parties are corrupted\cite{veugen2015secure}. This is because the only values seen by the parties during the computation are secret shares (that reveal nothing about the values they hide). To achieve security in the presence of malicious adversaries who will deviate from the protocol's specification, it is necessary to rely on specific methods to prevent cheating \cite{ishai2009secure}.

\subsection{Brute Force attack analysis}
Secure sharing of the algorithm parameters is necessary but not sufficient to guarantee privacy through our Hash-Comb encoding. A malicious observer could still crack our algorithm using a \emph{brute-force} approach to guess the range $[x_{min},x_{max}]$, hash the values and match the results with the target to validate the guess. It should be noted that the following considerations are intended for normalized values of the training datasets as well as of the learning model parameters.

Because the time complexity of brute force is $O(n\cdot m)$, where $n$ characters in a string of length $m$ requires $n*m$ tries, the smallest is the range of values and the more susceptible it is to brute-force attack. The authors in~\cite{TEZCAN2022102402} explain that a 128-bit security is well beyond the exhaustive search power of current technology. We can consider a secret on 128-bit computationally secure as to simply iterate through all the possible values requires $2^{128}$ flips on a conventional processor \footnote{\url{https://en.wikipedia.org/wiki/Brute-force_attack}}, 
the Von Neumann-Landauer Limit can be applied to estimate the energy required as $\approx 10^{18}$ joules, which is equivalent to consuming 30 gigawatts of power for one year. This is equal to $30\cdot10^9W\cdot365\cdot24\cdot3600 s = 9.46\cdot10^{17} J$ or $262.7$ $TWh$ which is approximately 0.1\% of the yearly world energy production.

Being $h_{i}=f_{hash}([x_{min_{i}};x_{max_{i}}])$ the hashed value of the channel $i$ having range $[x_{min_{i}},x_{max_{i}}]$ and assuming the use of double-precision floating-point representation (according to the IEEE 754 standard), where a bit is dedicated for the sign, 11 bit for the exponent and 52 bits for the mantissa, the combination of the limits perfectly fits as a shared secret of 128 bits required for the encoding. Unfortunately, due to the nature of the scalars in our scenarios, which are either ML models' parameters (weights) or normalized features values, the potential attacker could make strong assumptions in guessing those ranges. 

For example, the attacker can assume model parameters having $x_{min} \in [-1, -0.5]$ and $x_{max} \in [0.5, 1]$, therefore the merging can be represented using $2^{52}*2^{52} = 2^{104}$ values. More conservative ranges such as $[-1, -0.25]$ and $[0.25, 1]$ only increase the possible representations to $2^{106}$, both assumptions significantly reduce the number of tries for the attacker. In the case of normalized values, the task for the attacker is further simplified by the fact that $x_{min}$ and $x_{max}$ are fixed and corresponding to the normalization range itself, and the only effort would be the guess of the enlargement value $\Delta$. 

In conclusion, we use an additional number of bits to encode $y$ concatenated to the other shared secret to achieve the target of 128-bits to calculate $h_{i}$ as $f_{hash}([x_{min_{i}};y;x_{max_{i}}])$ and to achieve a schema that is brute-force proof.

\section{Training Models with Hash-Combs} 
\label{THC}
We start our experimental validation by training a monolithic ML model using multi-quantized training data.
\subsection{Datasets}
\begin{figure}
  \includegraphics[width=\linewidth]{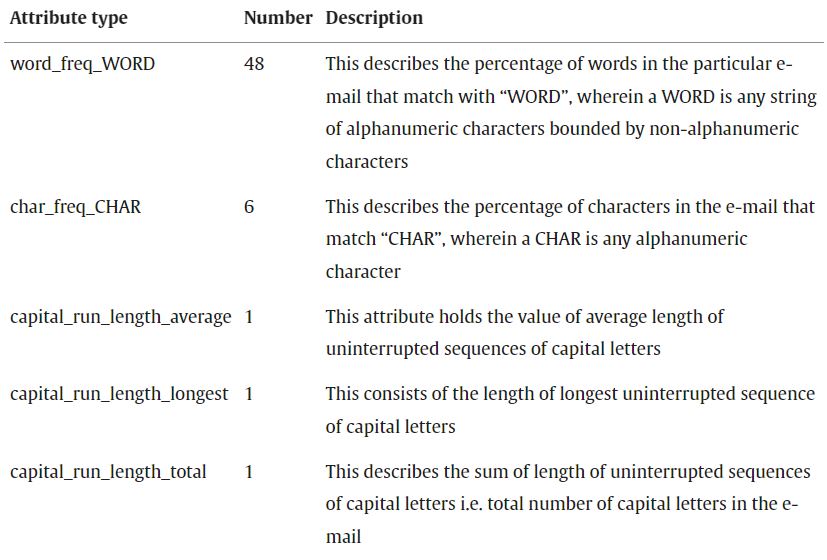}
  \caption{Some of the most relevant features in SPAM dataset}
  \label{fig:spamfeat}
\end{figure}

For our experiments, we used three different datasets. The first dataset is part of the UCI Machine Learning Repository \cite{misc_spambase_94}. Data consists of information from 4601 email messages, originally used in a study to identify junk email, or “spam”. For all 4601 email messages, the ground truth ("email" or "spam") is available, along with the relative frequencies of 57 of the most commonly occurring words and punctuation marks in the email message. Figure~\ref{fig:spamfeat} shows some of those features.
The second dataset, the recent IoT-23 dataset \cite{IoT23}, has been created simulating IoT network traffic generated by popular IoT devices, including a Philips HUE smart LED lamp, an Amazon Echo home intelligent personal assistant and a Somfy smart door lock. This dataset includes 23 traffic scenarios, 20 of which are malicious traffic from infected device and 3 are normal traffic. Such flows are labeled  "benign" or "malign", with an additional label specifying the attack type. For our experiment, we sampled the data from benign traffic and from malign traffic belonging to two types of attack, \textit{PartOfAHorizontalPortScan}  and \textit{Okiru}. Finally, the \textit{Cardiovascular Disease} \footnote{\url{https://www.kaggle.com/datasets/sulianova/cardiovascular-disease-dataset/data}} dataset consisting of 70000 records of patients data, 11 features divided as factual information, results of medical examination and information given by the patient. The target class represents the risk of a cardiovascular problem.  
\subsection{ML Model}
The ML model used for this experiments is a simple Multi-Layer Perceptron (MLP), implemented in Java 17. Our implementation can be configured to work as monolithic model as well as in federated learning with multiple instances of MLP communicating with a central server for sending/receiving updated weights and implementing \textit{FedAvg}. The model optimization technique is Stochastic Gradient Descent (SGD) and \emph{ReLu} is used as activation function. The architecture of our MLP counts 5 hidden layers, configured with $50, 25, 20, 25, 50$ neurons respectively. The learning rate is $\eta = 0.05$, while the number of epochs for the training depends on the requirements of the model to learn the specific dataset. The hardware is an Intel Core i5-9300H CPU 2.40GHz, 2400 Mhz, 4 Cores/8 Logical Cores. The same configuration for the model and hardware specification were maintained throughout all experiments.

\subsubsection{Monolithic Training}
The results in Table~\ref{tab:monolithicModel} conveys the performance of the monolithic training over the four datasets after normalization. The accuracy of the results was verified by comparison with the \textit{MLPClassifier}\footnote{\url{https://scikit-learn.org/stable/modules/generated/sklearn.neural_network.MLPClassifier.html}} availbable in \textit{Sklearn}, a Python library commonly used in similar research projects which has provided perfectly matching performance, therefore proving the accurateness of our implementation. The purpose of this test is to set a benchmark for the experiments in the following section. In fact, the comparison of the monolithic training with the results of the training paired with privacy preserving methods exemplify the performance improvement or decay of the latter when applied to FL.

\begin{table*}[]
\caption{Training and Testing on  a monolithic model}
\begin{adjustbox}{center}
\begin{tabular}{ll|ll|ll|}
\cline{3-6}
                                                &                                & \multicolumn{2}{c|}{\textbf{Training}}                                    & \multicolumn{2}{c|}{\textbf{Testing}}                                        \\ \hline
\multicolumn{1}{|l|}{\textit{\textbf{Dataset}}} & \textit{\textbf{Target label}} & \multicolumn{1}{l|}{\textit{\textbf{Support}}} & \textit{\textbf{Epochs}} & \multicolumn{1}{l|}{\textit{\textbf{Accuracy}}} & \textit{\textbf{F1-score}} \\ \hline
\multicolumn{1}{|l|}{\textit{Spam}}             & Spam                           & \multicolumn{1}{l|}{1813/3732}                 & 25K                      & \multicolumn{1}{l|}{0.8801}                     & 0.8804                     \\ \hline
\multicolumn{1}{|l|}{\textit{IoT23-Okiru}}      & Malicious                      & \multicolumn{1}{l|}{9890/20286}                & 10K                      & \multicolumn{1}{l|}{0.9911}                     & 0.9911                     \\ \hline
\multicolumn{1}{|l|}{\textit{IoT23-HPS}}        & Malicious                      & \multicolumn{1}{l|}{7478/18624}                & 40K                      & \multicolumn{1}{l|}{0.8280}                     & 0.8224                     \\ \hline
\multicolumn{1}{|l|}{\textit{Coronary}}   & Cardio                         & \multicolumn{1}{l|}{10465/21004}               & 12K                      & \multicolumn{1}{l|}{0.6231}                     & 0.6231                     \\ \hline
\end{tabular}
\end{adjustbox}
\label{tab:monolithicModel}
\end{table*}

\section{Experiments}
\label{E}
We experimented our Hash-Comb representation applied to  ML weights, encoded and communicated by local models to the central unit in a FL protocol. The experiments in Section~\ref{set1_2} aim at comparing different quantization levels of Hash-Comb in order to find the optimal number of quantizers and thus validate the theory as explained in Section~\ref{MHR}, while in Section~\ref{set_3} the focus is on comparing our approach to the most recent competing methods so as to demonstrate effectiveness through performance juxtaposing. 

Regarding the metric, we adopted the $F1$ score which computes an average of precision and recall. The $F1$ score is proven to be a valid performance metric when trying to solve binary classification problems as in our experiments. 

\begin{table*}[]
\caption{Testing results for Experiment 1}
\begin{adjustbox}{center}
\begin{tabular}{cc|ccccc|}
\cline{3-7}
\multicolumn{1}{l}{}                       & \multicolumn{1}{l|}{}                         & \multicolumn{5}{c|}{\textit{\textbf{F1-Score}}}                                                                                                                     \\ \cline{2-7} 
\multicolumn{1}{l|}{}                      & \multicolumn{1}{l|}{\textit{\textbf{Epochs}}} & \multicolumn{1}{l|}{$NoHC$}   & \multicolumn{1}{l|}{$HC_4$}    & \multicolumn{1}{l|}{$HC_6$}             & \multicolumn{1}{l|}{$HC_8$}             & \multicolumn{1}{l|}{$HC_{10}$} \\ \hline
\multicolumn{1}{|c|}{\textit{Spam}}        & 6K*4                                          & \multicolumn{1}{c|}{0.8856} & \multicolumn{1}{c|}{0.7019} & \multicolumn{1}{c|}{\textbf{0.9084}} & \multicolumn{1}{c|}{\textbf{0.9002}} & \textbf{0.8969}           \\ \hline
\multicolumn{1}{|c|}{\textit{IoT23-Okiru}} & 2.5K*4                                        & \multicolumn{1}{c|}{0.9938} & \multicolumn{1}{c|}{0.9870} & \multicolumn{1}{c|}{0.9935}          & \multicolumn{1}{c|}{\textbf{0.9941}} & \textbf{0.9941}           \\ \hline
\multicolumn{1}{|c|}{\textit{IoT23-HPS}}   & 12K*4                                         & \multicolumn{1}{c|}{0.8226} & \multicolumn{1}{c|}{0.6255} & \multicolumn{1}{c|}{0.7577}          & \multicolumn{1}{c|}{\textbf{0.8256}} & 0.8224                    \\ \hline
\multicolumn{1}{|c|}{\textit{Coronary}}    & 3K*4                                          & \multicolumn{1}{c|}{0.6217} & \multicolumn{1}{c|}{0.5960} & \multicolumn{1}{c|}{0.6157}          & \multicolumn{1}{c|}{\textbf{0.6229}} & \textbf{0.6223}           \\ \hline
\end{tabular}
\end{adjustbox}
\label{table:exp1}
\end{table*}

\begin{figure*}
  \includegraphics[width=\linewidth]{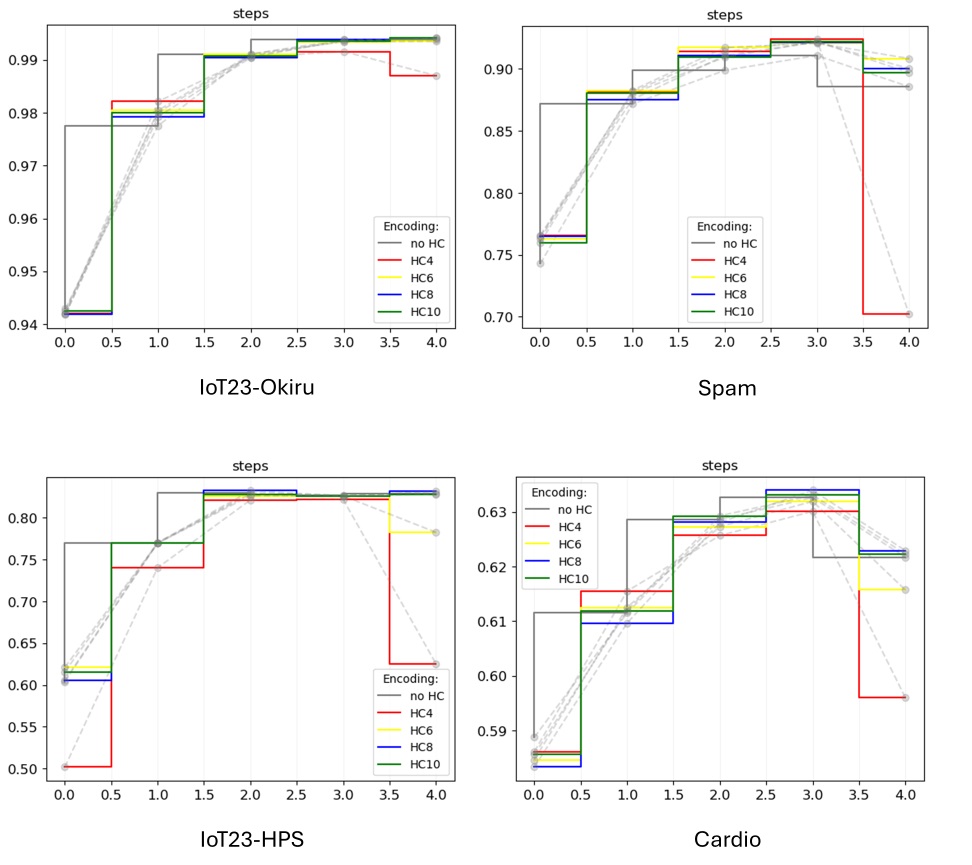}
  \caption{Training and Testing (last step) from Experiment 1}
  \label{fig:exp1}
\end{figure*}

\subsection{Federated Learning with Hash-Combed parameters }
\label{set1_2}

We replicated rhe monolithic training in Table~\ref{tab:monolithicModel} in a federated environment by deploying 4 instances of MLP communicating with the central unit for averaging. At each round, each local model is trained with exactly 25\% of randomly selected samples from the original set, the weights are sent to the central unit once a training step is completed. The purpose of the following 2 experiments is to verify the impact of Hash-Comb on the overall training and also to capture its performance by varying the number of quantizers. 

When the weights are sent in clear ($NoHC$), the standard \textit{FedAvg} Algorithm~(\ref{eq:FedAvg}) is applied on the central unit. While using Hash-Comb ($HC_L$), the quantization described in Section~\ref{MHR} is applied based on the initial range $S = [x_{min}, x_{max}]$ and a sequence of $L$ hashes is generated for each $w$ at every round. 
The encoding that we use for the experiments is defined as $h_L(w)=f_{hash}(ch_i)$ where $i\in[1, 2^L]$, it represents the hashing function applied to $ch_i$, the $i^{th}$ channel at level $L$ containing $w$. This encoding is used in~(\ref{eq:FedAvgHC}), implemented in the central unit to compute the encoded average of the received hashed values.

\subsubsection{Experiment 1: Bounded communication rounds}

The first experiment's aftermath is outlined in Table~\ref{table:exp1}. It appears immediately evident a slight improvement resulting from the mere exploitation of the FL protocol ($NoHC$) when compared with the monolithic training, an outcome that has been widely reported in the literature. 

Most importantly, the use of Hash-Comb distinctly reveals a significant improvement with $l \geq 6$ quantization levels, in particular when $l = 8$ the model enhances its performances across all datasets. Furthermore, the diagrams in Figure~\ref{fig:exp1} suggest a major over-fitting due to the excessive rounding of the weights when $l$ is lower than 8. In conclusion, these findings confirms $l=8$ as a \textit{magic number} for quantization as previously theorized in Section~\ref{MHR}.

         \begin {equation}        
            {w_{t+1} \leftarrow \sum_{n=1}^N {w^n_{t}\over N}}
            \label{eq:FedAvg}
         \end{equation} 
         \begin{equation}
            \quad  w_{t+1} \leftarrow \sum_{n=0}^N {c_i h_i\over N}
            \begin{cases}
                c_i=0  & \text{if n=0} \\
                c_i=c_i+1  & \text{if $n>0$ and $h_L(w^n_{t}) = h_i$} \\
            \end{cases}
           \label{eq:FedAvgHC}
         \end{equation} 

\subsubsection{Experiment 2: Incremental number of communication rounds}
\label{exp2}
The second set of experiments (see Table~\ref{table:exp2}) has the objective of evaluating the use of Hash-Comb by increasing the frequency of communication rounds and, consequently, the number of FedAvg iterations over the same total number of epochs as before. 

\begin{figure*}
  \includegraphics[width=\linewidth]{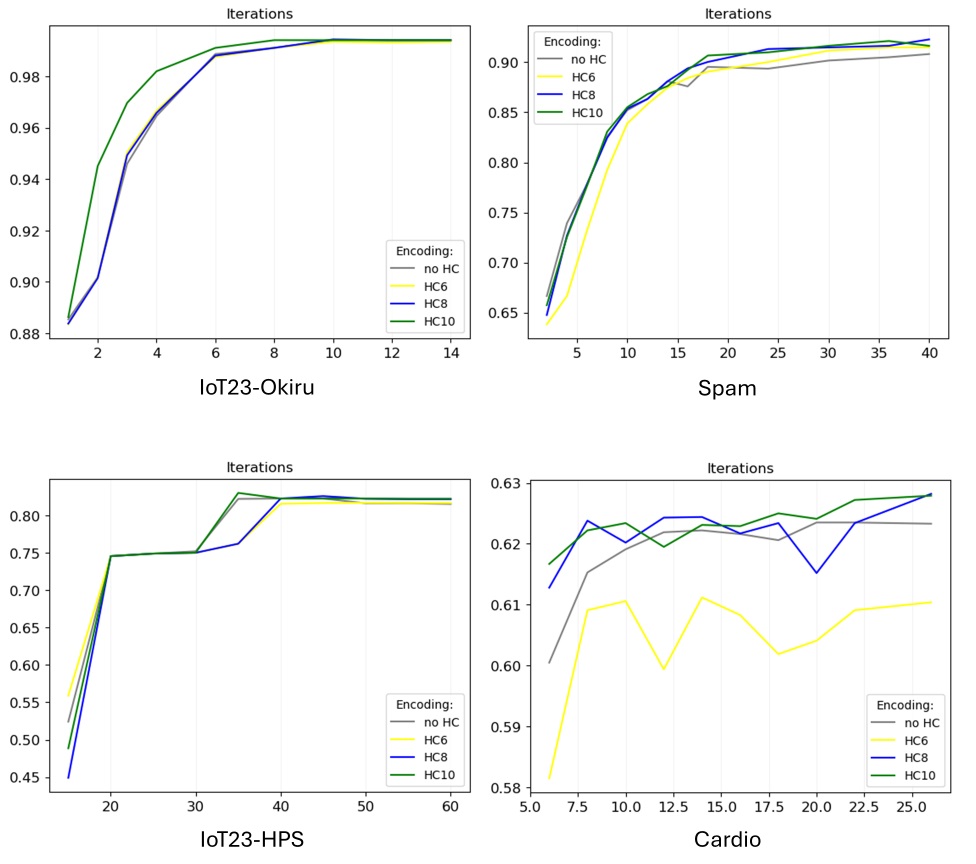}
  \caption{Model validation score at each FedAvg iteration from Experiment 2}
  \label{fig:exp2}
\end{figure*}

Each score within the diagrams in Figure~\ref{fig:exp2} is obtained by validating the general model at each round of communication, confirming the observations previously pointed out: the model using Hash-Comb with $L \geq 8$ achieves better results in the performance and sometimes is a faster learner when compared with non-quantized weights. Some minor instability is observed within the \textit{Coronary} dataset, probably due to insufficient data quality. In this case, averaging and rounding of the weights may cause fluctuating performance, noticeably for low $l$ values, though it eventually converges at the end of the training. This last observation suggests a possible correlation between the value $l$ and some measure of training data quality \cite{mauri2021estimating}.

\begin{table*}[]
\caption{Testing results for Experiment 2}
\begin{adjustbox}{center}
\begin{tabular}{cc|cccc|}
\cline{3-6}
\multicolumn{1}{l}{}                       & \multicolumn{1}{l|}{}                         & \multicolumn{4}{c|}{\textit{\textbf{F1-Score}}}                                                                                             \\ \cline{2-6} 
\multicolumn{1}{l|}{}                      & \multicolumn{1}{l|}{\textit{\textbf{Epochs}}} & \multicolumn{1}{c|}{$NoHC$}        & \multicolumn{1}{c|}{$HC_6$}         & \multicolumn{1}{c|}{$HC_8$}                  & \multicolumn{1}{c|}{$HC_{10}$} \\ \hline
\multicolumn{1}{|c|}{\textit{Spam}}        & 1K*40                                         & \multicolumn{1}{c|}{0.9082 (40)} & \multicolumn{1}{c|}{0.9148 (36)} & \multicolumn{1}{c|}{\textbf{0.9228 (40)}} & 0.9212 (36)               \\ \hline
\multicolumn{1}{|c|}{\textit{IoT23-Okiru}} & 1K*14                                         & \multicolumn{1}{c|}{0.9941 (10)} & \multicolumn{1}{c|}{0.9935 (10)} & \multicolumn{1}{c|}{\textbf{0.9944 (10)}} & 0.9941 (8)                \\ \hline
\multicolumn{1}{|c|}{\textit{IoT23-HPS}}   & 1K*60                                         & \multicolumn{1}{c|}{0.8224 (40)} & \multicolumn{1}{c|}{0.8164 (55)} & \multicolumn{1}{c|}{0.8221 (50)}          & \textbf{0.8299 (35)}      \\ \hline
\multicolumn{1}{|c|}{\textit{Coronary}}    & 1K*26                                         & \multicolumn{1}{c|}{0.6235 (20)} & \multicolumn{1}{c|}{0.6104 (26)} & \multicolumn{1}{c|}{\textbf{0.6282 (26)}} & {0.6279 (26)}      \\ \hline
\end{tabular}
\end{adjustbox}
\label{table:exp2}
\end{table*}

\subsection{Experiment 3: Comparison with classic DP }
\label{set_3}
In this Section our method is compared to a competing approach ~\cite{secFedAvg} functionally close to ours which is an ideal benchmark for our proposal. In fact, while the majority of related works focus on \textit{FedSGD} the method in the above mentioned work adds standard differential privacy (DP) to the \textit{FedAvg} algorithm. The authors claim that to achieve the desired $(\epsilon, \delta)$-DP, the required noise level is:

         \begin{equation}        
            {\sigma_{t,k}^2 = {2(\triangle f_{t,k})^2 ln(1.25q/\delta)\over \epsilon^2}}
            \label{eq:secFedAvg}
         \end{equation} 
         \begin{equation}
         {\triangle f \triangleq max_{D, D'}||f(D) - f(D') ||}
         \label{eq:deltaF}
         \end{equation}
         

The $\triangle f$ is so-called global sensitivity of function $f$ and being $D$ and $D'$ a neighboring datasets on the node \textit{k}, then:

         \begin{equation}
         {\triangle f_{t+1,k} = max||w^k_{t+1}(D_k) - w^k_{t+1}(D_k')||}
          \label{eq:globsens}
         \end{equation}

The equation in (\ref{eq:globsens})  represents the global sensitivity which is calculated as $\triangle f_{t+1,k}=4QG\eta_{t+1}$, where $Q$ is the number of SGD updates within each communication round, $q$ is the ratio between Q and the rows in the dataset and $G$ is the clipped local gradient on node $k$, whereas the learning rate is $\eta$. 
Based on the experiments mentioned in ~\cite{secFedAvg} and also on some preliminary tests we conducted on a training set with added Gaussian noise, the desirable variance is $\sigma^2 \approx 0.08$. In fact, a lower value does not guarantee the required privacy while a higher volume of noise significantly reduces the usability of the weights, resulting in a compromised learning outcome. 

In our experiment we aim at providing high protection level of $(2, 10^{-3})$-DP while retaining an acceptable usability of the weights. After plugging the above values into Formula~(\ref{eq:secFedAvg}), setting $G=2$ and a constant $\eta=0.05$, the desired $\sigma^2$ is obtained with $\textit{q}$ = ${SGD \over N_k}$ $\approx 0.008$, establishing a suitable number of gradient updates per communication round given the dataset size $N$.

\begin{figure*}
  \includegraphics[width=\linewidth]{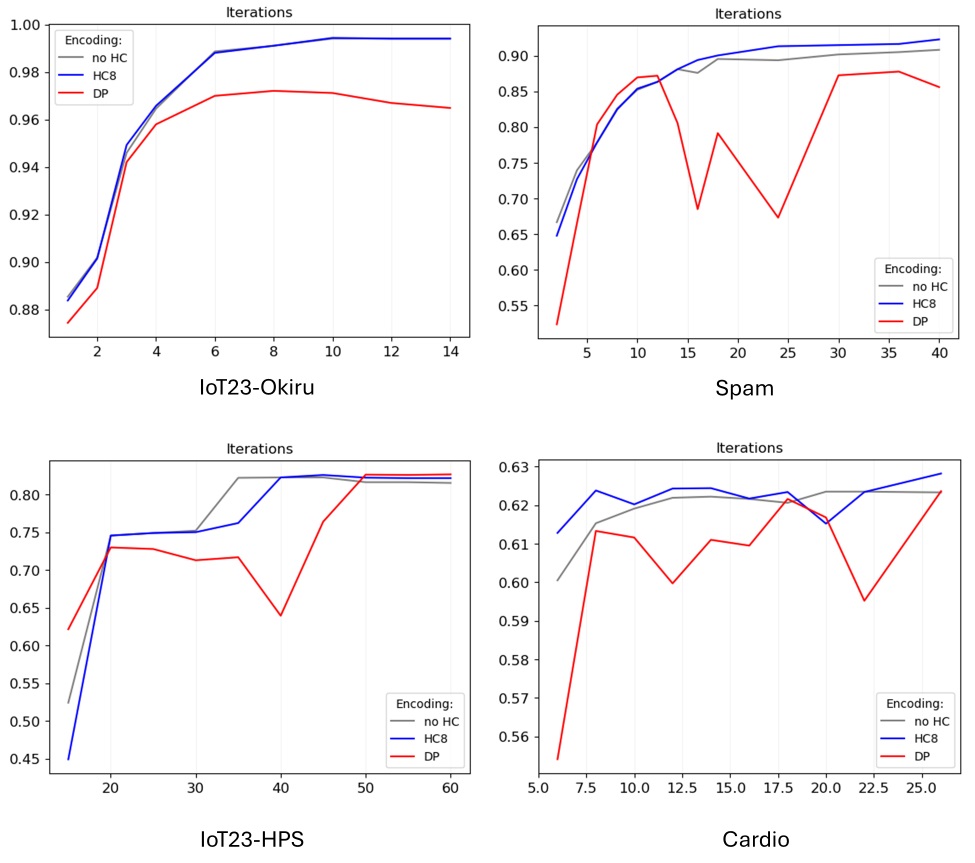}
  \caption{Comparison between Hash-Comb and DP}
  \label{fig:exp3}
\end{figure*}

The derived $(\epsilon, \delta)$-DP is applied to the training of our datasets for a direct comparison and the results are overlapped to ours and showed in Figure~\ref{fig:exp3}. From the analysis of the graphs it is immediately clear that despite a lower volume of noise, compensated by the high frequency of communications, the achieving of a desired DP level leads to a significant degradation of the model, both in terms of performance as well as learning convergence. 

Finally, the results in Table~\ref{table:exp3} show the performance gaining of our Hash-Comb ($HC_8$), featuring an average improvement  of $0.6\%$ in the F1-Score when compared to standard training and of $1.92\%$ when compared to $(\epsilon, \delta)$-DP.

\begin{table*}[]
\centering
\caption{Results of comparison between Hash-Comb and DP}
\begin{tabular}{cc|ccc|}
\cline{3-5}
\multicolumn{1}{l}{}                       & \multicolumn{1}{l|}{}                         & \multicolumn{3}{c|}{\textit{\textbf{F1-Score}}}                                                                                        \\ \cline{2-5} 
\multicolumn{1}{l|}{}                      & \multicolumn{1}{l|}{\textit{\textbf{Epochs}}} & \multicolumn{1}{c|}{$NoHC$}        & \multicolumn{1}{l|}{($\epsilon, \delta$)-DP} & \multicolumn{1}{l|}{$HC_8$} \\ \hline
\multicolumn{1}{|c|}{\textit{Spam}}        & 40K                                           & \multicolumn{1}{c|}{0.9082 (40)} & \multicolumn{1}{c|}{0.8777 (36)}                                         & \textbf{0.9228 (40)}     \\ \hline
\multicolumn{1}{|c|}{\textit{IoT23-Okiru}} & 14K                                           & \multicolumn{1}{c|}{0.9941 (10)} & \multicolumn{1}{c|}{0.9721 (8)}                                          & \textbf{0.9944 (10)}     \\ \hline
\multicolumn{1}{|c|}{\textit{IoT23-HPS}}   & 60K                                           & \multicolumn{1}{c|}{0.8224 (40)} & \multicolumn{1}{c|}{0.8264 (60)}                                         & 0.8221 (50)              \\ \hline
\multicolumn{1}{|c|}{\textit{Coronary}}    & 26K                                           & \multicolumn{1}{c|}{0.6235 (20)} & \multicolumn{1}{c|}{0.6236 (26)}                                         & \textbf{0.6282 (26)}     \\ \hline
\end{tabular}
 
 \label{table:exp3}
\end{table*}

\section{Conclusion and Outlook}
\label{C}
While data privacy is often mentioned as the motivation for Federated Learning schemes, many popular techniques for federated ML models training cannot yet achieve certifiable privacy and confidentiality. Also, they are often clumsy to implement.
We described a novel procedure that smoothly combines randomized quantization to achieve differential privacy and multi-hashing (using standard, regulation-compliant hashes) to provide privacy, confidentiality and regulatory compliance. We empirically studied the performance of our technique and demonstrated that, compared to classic Differential Privacy approaches, it delivers an improved privacy-accuracy trade-off and a much smaller footprint.

\section*{Acknowledgement}
We would like to express our sincere appreciation to ASPIRE for their financial support (Award Number AARE21-366).






\end{document}